\documentclass[twocolumn,showpacs,superscriptaddress,pre]{revtex4}
\usepackage{latexsym}
\usepackage{graphicx}
\usepackage{amssymb}

\newtheorem{theo}{\bf Theorem}

\def\R{\mathbb R}

\begin{document}

\title{Detectability of non-differentiable generalized synchrony}



\author{ Nikolai F. Rulkov}
\affiliation{Institute for Nonlinear Science, University of California,
San Diego, La Jolla, CA 92093-0402}
\author{Valentin S. Afraimovich}
\affiliation{ IICO-UASLP, A. Obreg\'on 64, 78000 San Luis Potos\'{\i},
SLP, M\'exico}

\begin{abstract}
Generalized synchronization of chaos is a type of cooperative
behavior in directionally-coupled oscillators that is characterized
by existence of stable and persistent functional dependence of
response trajectories from the chaotic trajectory of driving
oscillator. In many practical cases this function is
non-differentiable and has a very complex shape. The generalized
synchrony in such cases seems to be undetectable, and only the
cases, in which a differentiable synchronization function exists,
are considered to make sense in practice. We show that this
viewpoint is not always correct and the non-differentiable
generalized synchrony can be revealed in many practical cases.
Conditions for detection of generalized synchrony are derived
analytically, and illustrated numerically with a simple example of
non-differentiable generalized synchronization.

\end{abstract}

\maketitle

\narrowtext
\vskip1pc
\newcounter{eq}
\section{Introduction}\label{introduction}
Synchronization plays an important role both for understanding of
cooperative behavior in natural networks of
oscillators~\cite{Glass} and for various engineering
applications~\cite{Minorsky,Blekhman}. Recently a significant
interest in understanding and theoretical description of
synchronization regimes among the oscillators with chaotic behavior
is perceived, see for example recent books and
reviews~\cite{Pikovsky,Mosekilde,Boccaletti}. Various types of
chaos synchrony, whose description may require different
theoretical frameworks, were found in natural systems and
specified. These types of synchrony include identical
synchronization~\cite{AVR86,FY84,Pecora90}, generalized
synchronization~\cite{Rulkov95,Pecora95,Parlitz96,Rulkov2001} and
phase synchronization~\cite{Pikovsky97,Zaks99}.

The framework of generalized synchronization was proposed as an
attempt to extend the classical theory of forced synchronization of
a periodic oscillator, initiated by the works of van der
Pol~\cite{VDP}, Andronov and Witt~\cite{Andronov}, to the case of
directionally coupled chaotic oscillators. This framework defines
synchronization as the onset of conditional stability of a
chaotically driven oscillator, and as the existence of a functional
relation that maps the chaotic trajectory of driving oscillator
into the trajectory of driven
oscillator~\cite{Rulkov95,Abarbanel96}. In the case of invertible
dynamics of driving system such functional relation is usually
substituted with a function that maps the state of driving system
into the state of response one when these states are measured
simultaneously. Rigorous mathematical results indicate that,
depending on the strength of conditional stability, the
synchronization function can be differentiable or
non-differentiable~\cite{Stark97,Hunt97,Josic98,Stark99,Afr01}. In
many experimental studies researcher needs to establish the fact of
chaos synchronization when direct analysis of conditional stability
is hardly possible. In such a situation the detection of
generalized chaos synchrony characterized by a non-differentiable
function, which due to dense wrinkles, cusps and finite number of
points appears as a thick and fuzzy set, may seem to be
impossible~\cite{Hunt97,So02,Barreto}.

In this paper we show that detectability of the non-differentiable
synchrony can be significantly improved and become feasible if one
explores synchronization function taking into account additional
points on sufficiently long intervals of the driving trajectory
preceding to the current state. The paper is organized as follows.
In Section~\ref{Sec2} we discuss the idea of such detectability and
evaluate the improvement using numerical analysis of a simple
example. Section~\ref{Sec3} develops a theoretical arguments
explaining the mechanism behind the detectability improvement.
Section~\ref{Sec4} discusses possible effects caused by small
additive noise in the data. Summary of the results and possible
applications are discussed in the Conclusion.

\section{Numerical example}\label{Sec2}

To illustrate the idea of detectability enhancement we first
consider an example of a drive-response system which was proposed
and studied in~\cite{Hunt97}. In this example generalized baker map
\setcounter{eq}{\value{equation}}
\addtocounter{eq}{+1}
\setcounter{equation}{0}
\renewcommand{\theequation}{\theeq \alph{equation}}
\begin{equation}\label{mapx1}
x_{n+1}^{(1)}=\cases {
 \lambda_a x_n^{(1)},    &if  $ x_n^{(2)} < a,$ \cr
 \lambda_a+\lambda_b x_n^{(1)}, &if $ x_n^{(2)} \geq a,$ \cr}
\end{equation}
\begin{equation}
x_{n+1}^{(2)}=\cases {
 x_n^{(2)}/a,    &if $ x_n^{(2)} < a,$ \cr
 (x_n^{(2)}-a)/b, &if $ x_n^{(2)} \geq a,$ \cr
            } \label{mapx2}
\label{func}
\end{equation}
\newcommand{\refmap}{1}
where $0 \leq x_n^{(i)} <1$, $\lambda_a=1-\lambda_b=0.3$, and
$a=1-b=0.5$, drives a system of the form
\setcounter{equation}{\value{eq}}
\renewcommand{\theequation}{\arabic{equation}}
\begin{eqnarray}
y_{n+1}=c y_n+\cos(2\pi x_n^{(1)}). \label{mapy}
\end{eqnarray}
Here parameter $c$ defines the properties of the response behavior.
Consider the systems dynamics within the parameter interval
$0<c<1$. In this case response system (\ref{mapy}) is conditionally
stable. The dynamics of driving system (\refmap) is invertible and
according to the theory (see, for example~\cite{Stark99,Afr01})
there exists a continuous function $y_n=h({\bf x}_n)$, where ${\bf
x}_n=(x_n^{(1)},x_n^{(2)})$. Due to the specific form of the
driving and response systems function $h$, in our case, is
independent of $x_n^{(2)}$. Indeed, given the value of $x_n^{(1)}$
all previous values of this variable can be found from equation
(\ref{mapx1}), when one iterates this one-dimensional map backward
in time, and these values are independent of $x_n^{(2)}$.
Therefore, function $h$ can be plotted as a graph in the variables
plane $(x_n^{(1)},y_n)$.

\begin{figure}
\begin{center}
\includegraphics[width=7.5cm]{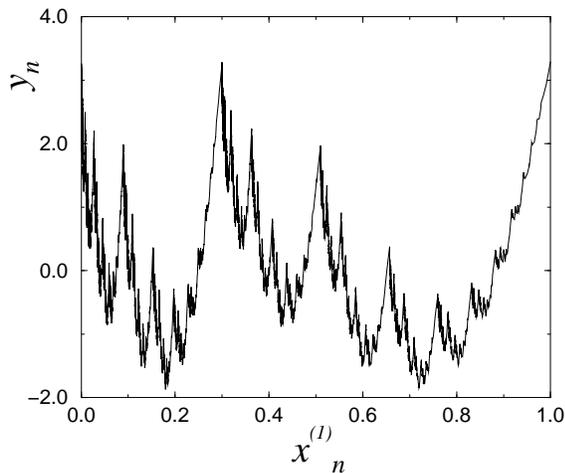}
\end{center}
\caption{The shape of the function $y_n=h(x^{(1)}_n)$
computed with $c=0.7$.}
\label{fig1}
\end{figure}

The example of non-differentiable function $h$ computed with
$c=0.7$ is shown in Fig.~\ref{fig1}. It is clear from the shape of
the function that, in practical situation with a similar function,
the existence of the function cannot be revealed from such a plot
because the states of response system measured for nearby states of
driving system can be very disperse. This situation can make one to
believe that the onset of non-differentiable generalized
synchronization is practically undetectable. The statements on such
practical undetectability are usually made when one analyses only
relation between simultaneous states in the attractors of driving
and response systems. One may ask a question if additional
information about the chaotic trajectory can help to resolve the
complexity of this functional relation? And if the answer is yes,
what properties of the non-differentiable function can be improved?

To illustrate positive answer to the first question we adopt the
approach developed in ~\cite{ACR02} and consider the additional
information about the driving chaotic trajectory ${\bf x}_n$ using
its symbolic description. Now we define the state of the driving
system as the value of $x^{(1)}_{n-m}$ and the symbolic sequence
$[\alpha_1,...,\alpha_m]$ generated in the next consecutive
iterations towards $x^{(1)}_{n}$. In the considered example symbols
$\alpha_i$ can be easily defined from the evolution of variable
$x^{(2)}_n$. If $x^{(2)}_{n-i}<a$, then $\alpha_i=0$. If
$x^{(2)}_{n-i}\geq a$, then $\alpha_i=1$. From the data generated
by the maps (\refmap), (\ref{mapy}) we can examine synchronization
function in a new form $h^{(m)}$ which is defined as a mapping
$([\alpha_1,...,\alpha_m],x^{(1)}_{n-m}) \rightarrow y_n$.

\begin{figure}
\begin{center}
\includegraphics[width=7.5cm]{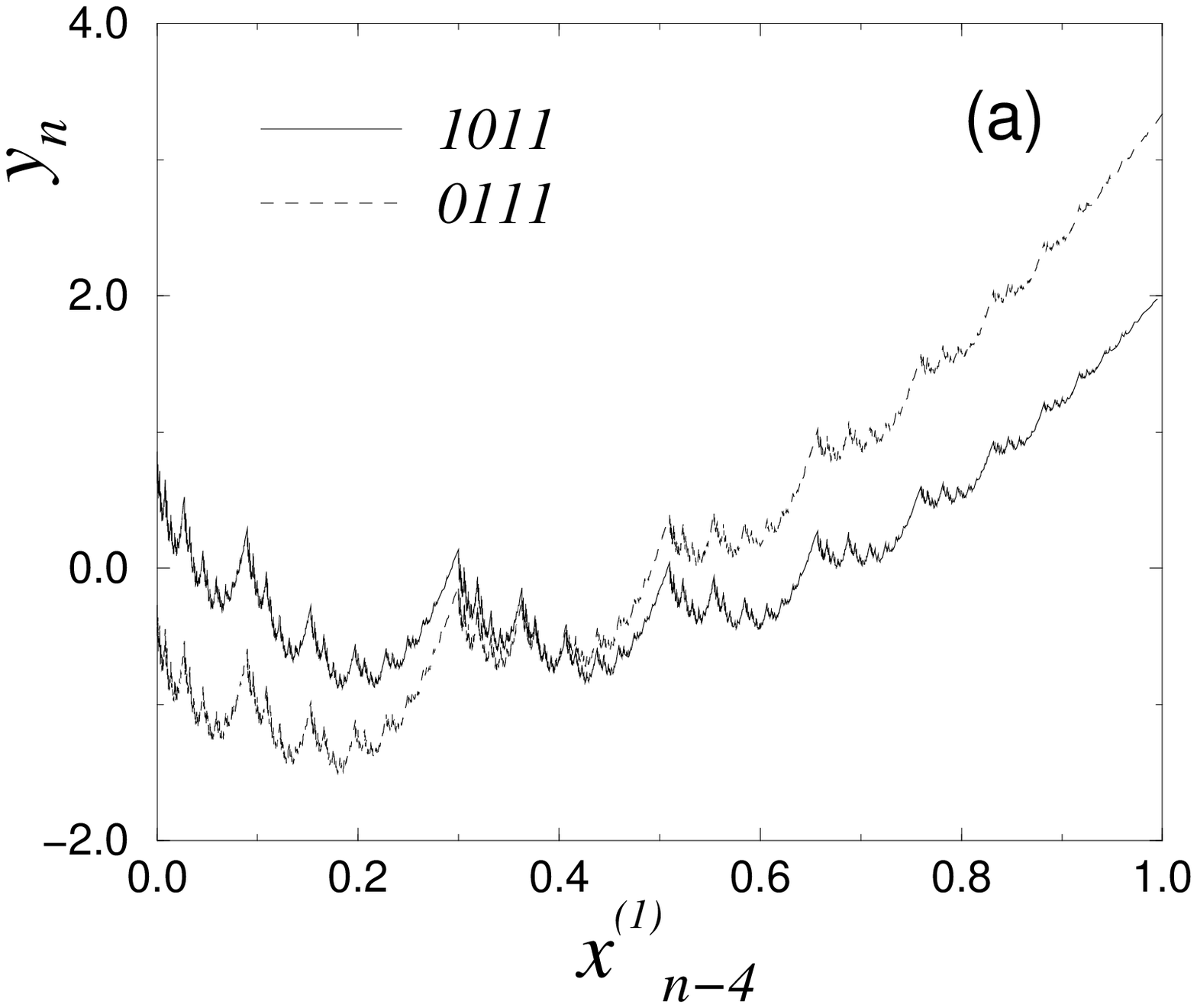}
\includegraphics[width=7.5cm]{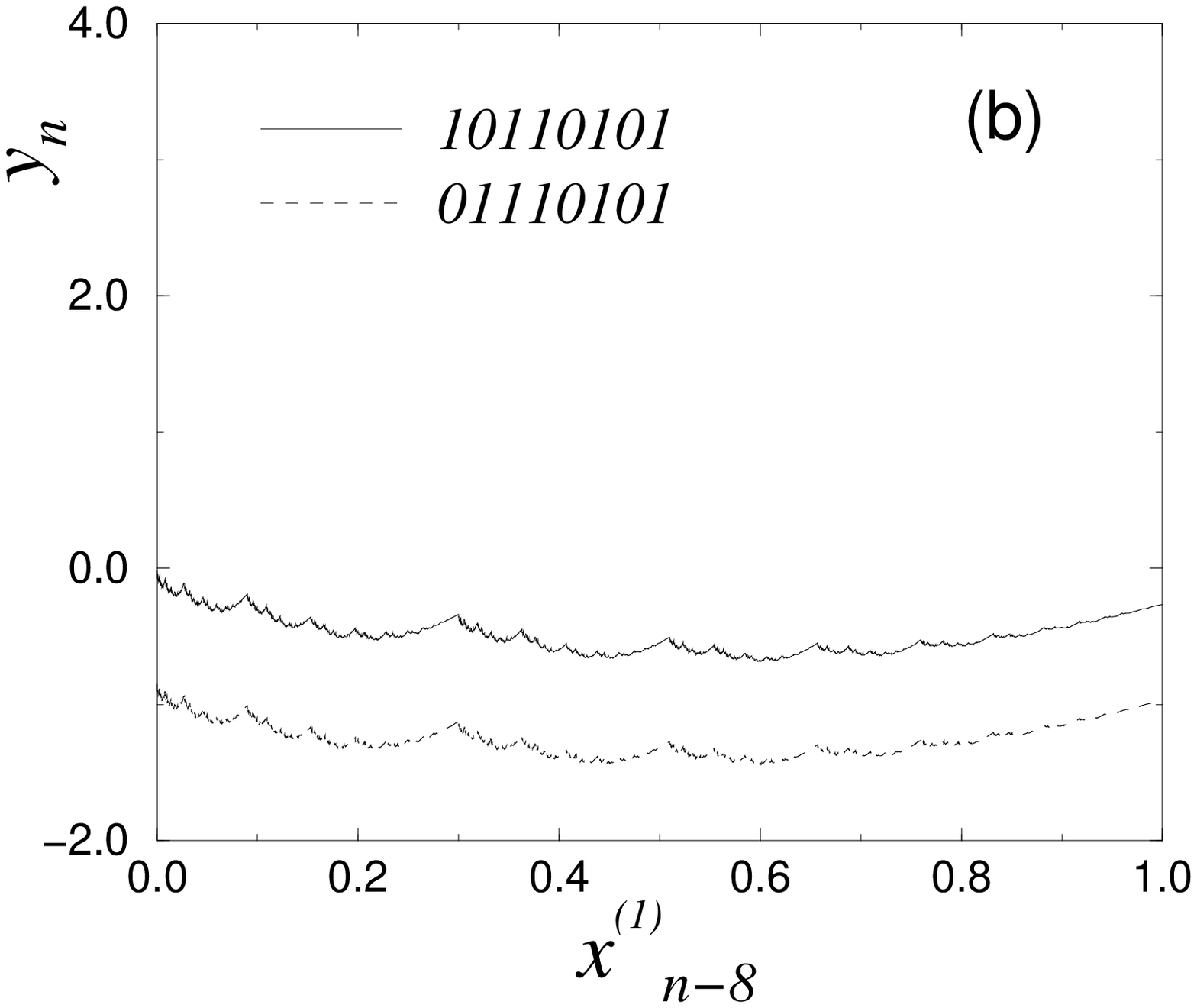}
\end{center}
\caption{The synchronization function computed using additional symbolic
information on the driving trajectory for the case shown in
Fig~\ref{fig1}. Only two symbolic masks of each length $m$ are
presented. The case $m=4$ is shown in panel (a) and $m=8$ is in
panel (b). }
\label{fig2}
\end{figure}

In order to illustrate the improvement of the modified
synchronization function $h^{(m)}$ with the increase of $m$ we plot
$y_n$ versus $([\alpha_1,...,\alpha_m],x^{(1)}_{n-m})$ for two
fixed symbolic sequences that differ by two most recent symbols.
The cases of $m=4$ and $m=8$ are presented in Fig.~\ref{fig2}a and
b respectively, where the parameters of the maps are the same as in
Fig.~\ref{fig1}. Comparing these plots with the plot shown in
Fig.~\ref{fig1} one can see that the existence of synchronization
function becomes more apparent as the delay $m$ increases. Notice,
that the scales of corresponding axes in these plots are the same.

\begin{figure}
\begin{center}
\includegraphics[width=7.5cm]{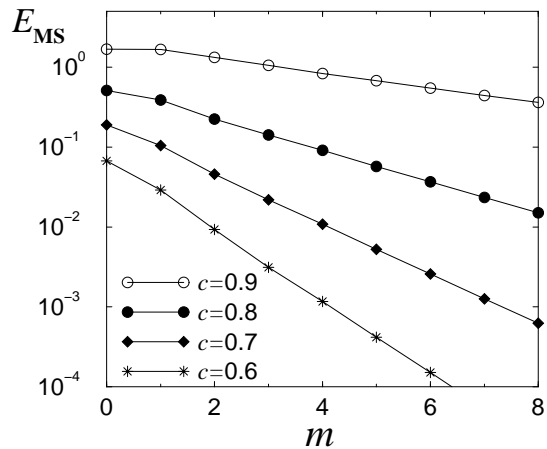}
\end{center}
\caption{The dependence of mean squared error of best polynomial
fitting function for the attractor points $(x_{n-m},y_n)$ on the
length $m$ of the preceding masks $S^i_m$.}
\label{fig3}
\end{figure}

We studied how the complex image of synchronized attractors in the
space of a drive-response system converges to a "good" simple
function with the increase of $m$. We analyzed the sets of
attractor points conditioned by all possible symbolic masks
$\alpha$ of various lengths $m$. For each mask of preceding symbols
$S^i_m=[\alpha_1,..,\alpha_m]$ we computed the best polynomial
fitting function $\phi_{S^i_m}(x)$ of order 30 using Singular Value
Decomposition (SVD) algorithm, and studied the dependence of mean
squared error ($E_{MS}$), averaged over all masks of length $m$, on
the value of $m$. This dependence computed for four different
values of $c$ is shown in Fig.~\ref{fig3}.

\begin{figure}
\begin{center}
\includegraphics[width=7.5cm]{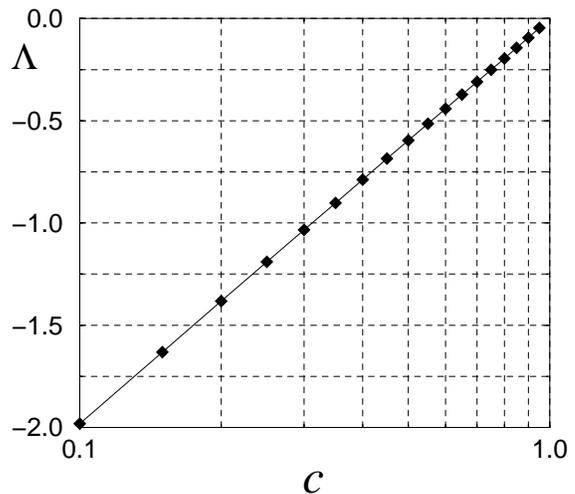}
\end{center}
\caption{The dependence of convergence rate $\Lambda$
on the value of coupling parameter $c$ plotted in the logarithmic
scale.}
\label{fig4}
\end{figure}

One can see from Fig.~\ref{fig3} that $E_{MS}$ decreases
exponentially fast when $m$ increases. Approximating this
dependence with exponential,
\begin{eqnarray}
 E_{MS}(m)&\sim&e^{\Lambda m},\label{mse}
\end{eqnarray}
one can find the rate of convergence $\Lambda$. Figure~\ref{fig4}
shows how the convergence rate $\Lambda$ evolves with the change of
parameter value $c$. The absolute value of $\Lambda$ decreases as
the value of $c$ grows. This indicates that for higher values of
$c$ the synchronization function becomes more complex~\cite{Hunt97}
and its detection with a given resolution requires more information
on the driving trajectory then for lower values of $c$.

One can easily check that when $c \to 1$, the conditional dynamics
of response system (\ref{mapy}) approaches the threshold of
instability and, as the result, synchronization terminates and
function $h$ disappears. The plot of $\Lambda$ vs $c$ reflects this
fact and one can see from Fig.~\ref{fig4} that convergence rate
$\Lambda$ tends to zero as $c \to 1$. It can be shown that the
linear dependence of $\Lambda$ on $\log(c)$ is due to the fact that
the response system in this example is a linear system and
$\log(|c|)$ is the contraction rate of its phase volume.

\begin{figure}[t]
\begin{center}
\includegraphics[width=6.5cm]{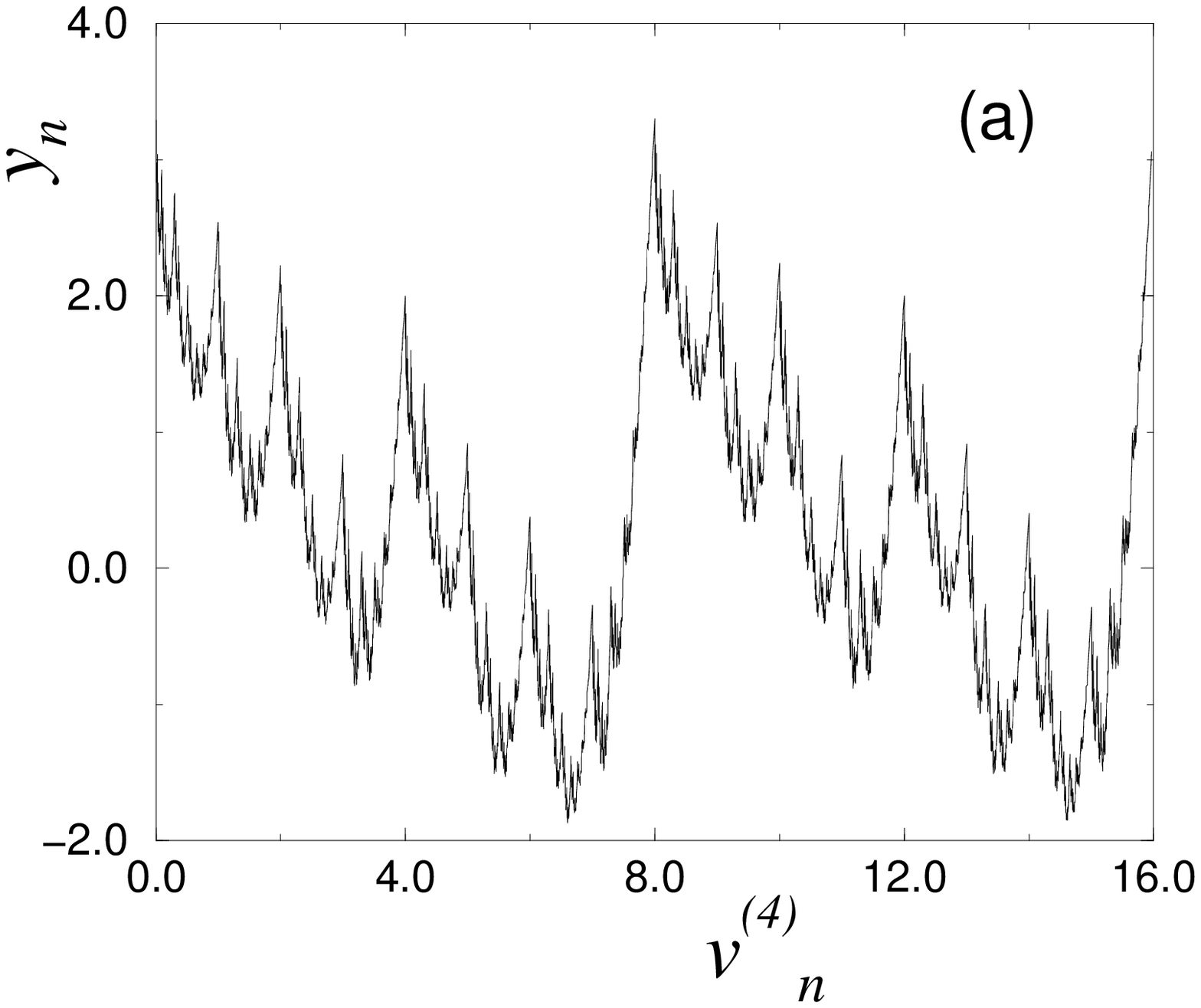}
\includegraphics[width=6.5cm]{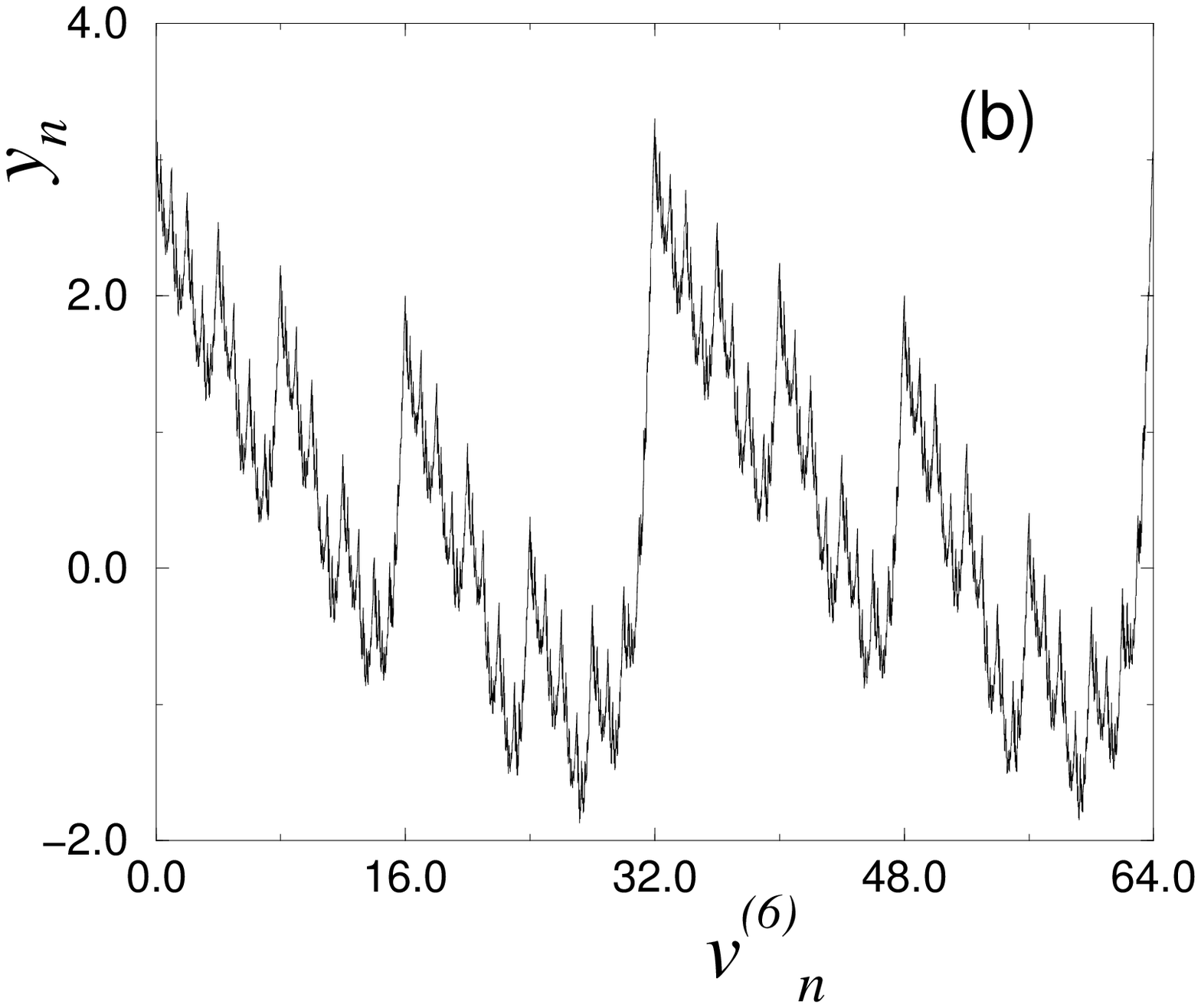}
\includegraphics[width=6.5cm]{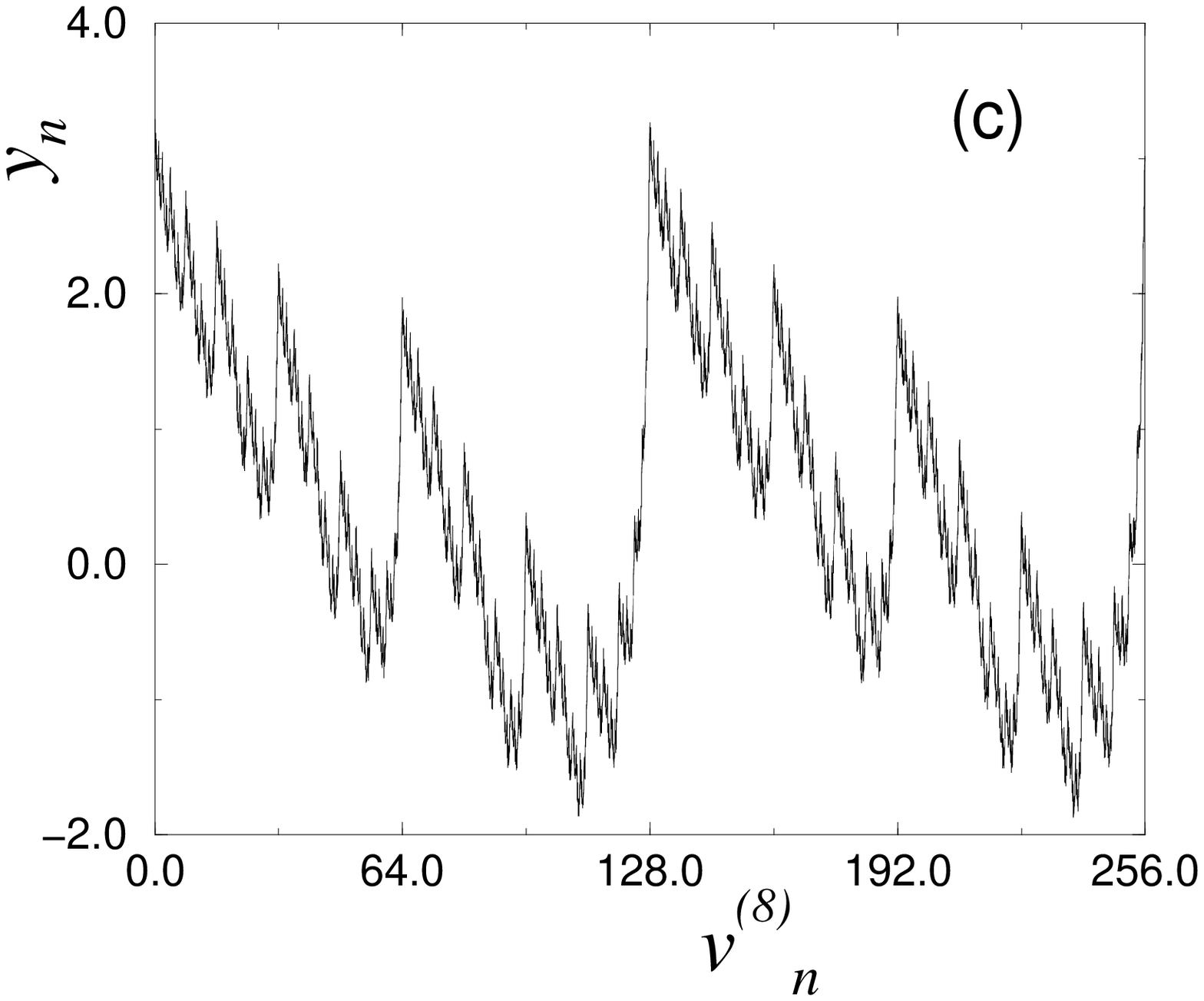}
\end{center}
\caption{The synchronization function shown in Fig.~\ref{fig1} plotted
versus a new variable $v^{(m)}_n:=I(S^i_m).x^{(1)}_{n-m}$ that
contains symbolic information about the driving trajectory. Panel
(a) shows the case $m=4$, (b) $m=6$, and (c) $m=8$. Note the
changes in the values of the horizontal axes.}
\label{fig5}
\end{figure}

To get a better view on the function improvement, examine how
additional symbolic information collected along the driving
trajectory changes the shape of the whole synchronization function.
One way of taking such symbolic information into account is to
compute the integer value out of binary symbolic mask $S^i_m$ and,
then, supplement this integer with the fractional value given by
$x^{(1)}_{n-m}$. Note that in our case $0 \leq x^{(1)}_{n-m}<1$.
Computing the integer part we assume that the most resent symbol
$\alpha_1$ of mask $S^i_m$ is the most significant bit. As the
result we obtain the decimal values of the form
$I(S^i_m).x^{(1)}_{n-m}$. Every decimal value is considered as a
new argument of the modified synchronization function $h^{(m)}$.
Figure~\ref{fig5} presents such a function plotted for three
different values of $m$. One can see that for large $m$ ($m>4$),
the overall shape and complexity of the function remains about the
same, but the interval of the argument increases in size by factor
$2^m$. This indicates that there exists some kind of
self-similarity of the non-differentiable synchronization function,
and the enhanced detectability is the result of more precise
evaluation of the state of the driving system. It is important to
note that the precision of the state evaluation increases with $m$
despite the fact the values of variable $x^{(1)}$ are measured with
the same precision as before.

\section{Theoretical results}\label{Sec3}

The example considered above clearly indicates that the functional
relation between the synchronized system becomes more apparent when
length of driving trajectory taken for the analysis increases. In
order to examine which properties of the non-differentiable
function change and simplify the detection of the functional
relation we present the following theoretical analysis.

In this section we shall concentrate on  systems with a
unidirectional coupling (or systems with a skew product structure)
of the form:
\begin{equation}
\begin{array}{rcl}\label{definition}
x'&=&f(x)\\ y'&=&g_{\rho}(x,y)\,.
\end{array}
\end{equation}
These equations determine a map $F_{\rho}:(x,y)\mapsto (x',y')$
generating a dynamical system. The first subsystem is called the
driving system, the second subsystem is called the response system
and ${\rho}$ is a parameter that controls the coupling strength.
The fact of synchronization in this systems means that there is a
region of parameter values ${\rho}$ in which, for any initial
conditions $(x_0,y_0)$, $(x_0,\tilde{y}_0)$,
\begin{equation}\label{distgoestozero}
\lim_{n\to\infty}dist (y_n,\tilde{y}_n)=0,
\end{equation}
where $(x_n,y_n)=F_{\rho}^n(x_0,y_0)$
($(x_n,\tilde{y}_n)=F_{\rho}^n(x_0,\tilde{y}_0)$). Loosely
speaking, this means that, for any initial conditions, the distance
between the states of the slave subsystem goes to zero with time.

We assume for the sake of definiteness, that in the system
(\ref{definition}) one has $x\in\R^d$ and $y\in\R^\ell$, and that
$g_{\rho}$ is continuous and $f$ is a homeomorphism. Since we study
dissipative systems we also assume that there exists a ball of
dissipation $B\subset\R^{d+\ell}$, i.e. $F_{\rho}(B)\subset Int(B)$
for any ${\rho}\in S$, where $S$ is a region in ${\rho}$-space (in
which the system (\ref{definition}) exhibits synchronization).
Without loss of generality we assume that $B=B_x\times B_y$, i.e.
$B$ is a rectangle, where $B_x$ (resp. $B_y$) is a ball in x-space
(resp. y-space). Denote by ${\cal A}_{\rho}$ the maximal attractor
in $B$, i.e. ${\cal A}_{{\rho}}=\cap_{n=0}^{\infty}F^n_{\rho}(B)$.

Through this section we shall assume that one-to-one globally
stable generalized synchronization occurs in $B$, i.e. the
condition (\ref{distgoestozero}) satisfy when $(x_0,y_0)$ and
$(x_0,\tilde{y}_0)$ are arbitrary points in $B$.

It was shown in~\cite{Afr01} that under these conditions there is a
continuous functional dependence between $x$ and $y$--components of
orbits while the system is in synchronized region. To obtain more
detailed characteristics about this functional dependence we need
an additional assumption. Assume that
\begin{equation}\label{ms2}
|y_{n+1}-\tilde{y}_{n+1}|\leq c | y_n-\tilde{y}_n|
\end{equation}
where $c<1$. Of course the parameter $c$ is a function of ${\rho}$.
For the sake of simplicity, we assume that $c={\rho}$. Thus,
\begin{equation}\label{ms3}
|y_{n+1}-\tilde{y}_{n+1}|\leq {\rho} | y_n-\tilde{y}_n|,\quad
0<{\rho}<1.
\end{equation}
It follows that
\begin{equation}\label{lip0}
|g_{\rho}(x,y)-g_{\rho}(x,\tilde{y})|\leq {\rho} |y-\tilde{y}|
\end{equation}
for any $(x,y)$, $(x,\tilde{y})\in B$. Let us draw the attention on
the fact that the smaller ${\rho}$ is, the greater the coupling
strength.

Assumption (\ref{ms3}) implies that $| y_n-\tilde{y}_n|$ goes to
zero exponentially fast, and this fact allows one to prove that
function $h: x_n \mapsto y_n$ is H\"older continuous provided that
the functions $f$ and $g_{\rho}$ have good smooth properties, or at
least they are Lipschitz-continuous. So we assume that:
\begin{equation}\label{lip1}
|f(x)-f(\tilde{x})|\leq \gamma_{+} |x-\tilde{x}|
\end{equation}
and
\begin{equation}\label{lip2}
|f^{-1}(x)-f^{-1}(\tilde{x})|\leq \gamma_{-} |x-\tilde{x}|\,,
\end{equation}
where $\gamma_{-},\gamma_{+}\geq 1$. Here $\gamma_{+}$
characterizes the rate of divergence of nearby driving trajectories
forward in time, and $\gamma_{-}$ characterizes their divergence
backward in time. Moreover, we assume that:
\begin{equation}\label{lip3}
|g_{\rho}(x,y)-g_{\rho}(\tilde{x},y)|\leq \eta |x-\tilde{x}|
\end{equation}
for any $(x,y)$, $(\tilde{x},y)\in B$, where $\eta>0$.

The following statement was proved in~\cite{Afr01}.

\begin{theo}[H\"older property]\label{holder}
Under  assumptions  (\ref{lip0})-(\ref{lip3}) the function $h$ is
H\"older continuous, i.e. for any $0<\alpha<\alpha_{0}$,
$x,\tilde{x}\in {\cal A}_{{\rho},x}$ one has:
\begin{equation}\label{holder1}
|h(x)-h(\tilde{x})|\leq 2a |x-\tilde{x}|^{\alpha}
\end{equation}
where
\begin{equation}\label{holder2}\alpha\leq \alpha_0\equiv
\left(1-\frac{\log(\gamma_{+}\gamma_{-})}{\log {\rho}}\right)^{-1}\;,
\end{equation}
and $a\geq a_0$, where $a=a_0$ is the solution of the equation:
\[
a=\frac{\eta}{\gamma_{+}-{\rho}}\,a^{\frac{\log
(\gamma_{+}\gamma_{-})}{\log {\rho}}}
\left(\gamma_{+}\gamma_{-}\right)^{1-
\frac{\log\vert B_{y}\vert}{\log {\rho}}}\,\cdot
\]
\end{theo}

Here $|B_{y}|$ stands for the diameter of $B_{y}$. Recall that
H\"older exponents quantify the ``degree of
non-differentiability''.

Our goal is to understand what happens if one tries to study the
dependence between the $y$--coordinate of the orbit at iteration
$n$ and the $x$--coordinate at the moment $n-m$, for $m>0$. In
other words we are going to study the effect of the ``delay'' onto
the functional dependence in the synchronized region. The following
result holds:

\begin{theo}\label{theorm2}
Let conditions of Theorem~\ref{holder} be satisfied and
$(x_n,y_n)$, $(x'_n,y'_n)$ be orbits belonging to ${\cal
A}_{\rho}$. Then for every $\epsilon>0$ there exists $\delta>0$
such that for every pair $x_{n-m},\tilde{x}_{n-m}$,
$|x_{n-m}-\tilde{x}_{n-m}|<\delta$ one has
\begin{equation}\label{estimate}
|y_n-\tilde{y}_n| \leq A |x_{n-m}-\tilde{x}_{n-m}|^{\alpha}
\end{equation}
where
\begin{equation}\alpha\leq \alpha_0+ \beta(\epsilon),
\;\;\;
\alpha_0
\equiv
\left(1-\frac{\log(\gamma_{+}\gamma_{-})}{\log {\rho}}\right)^{-1}\;,
\end{equation}
and $A\geq A_m$, where
\begin{equation}\label{m-constant}
A_m=\epsilon + \varrho \gamma_{-}^{-m}
\end{equation}
and $\varrho$ is a constant independent of $m$.
\end{theo}

{\em Proof}. Without loss of generality we prove the
estimate~(\ref{estimate}) for $n=0$. The proof  for another values
of $n$ is the same. Let ${\cal A}_{{\rho},x}\equiv\Pi_x{\cal
A}_{\rho}$ be the image of ${\cal A}_{\rho}$ under the natural
projection $\Pi_x$ to $\R^d$.

Consider a point $x_0\in {\cal A}_{{\rho},x}$. Let $x_{-i}\equiv
f^{-i}(x_{0})$. Given the backward orbit $\{x_{-i}\}_{i=0}^\infty$,
the dynamics on $B_{y}$ is defined by the sequence of operators
$\{g_{\rho}(x_{-i},\cdot)\}_{i=0}^\infty$ acting on $B_{y}$. Define
the following operation:
$$
(g_{\rho}\star g_{\rho})(x,y)\equiv g_{\rho}(f(x),g_{\rho}(x,y))\,.
$$
We denote by $g_{\rho}^{\star k}(x,y)$ the result of the operation
`$\star$' performed $k$ times (by convention $g_{\rho}^{\star
0}\equiv g_{\rho}$). Notice that $g_{\rho}^{\star k}(x,y)=\Pi_{y}
F_{\rho}^{k}(x,y)$.

Consider two points $(x_0,y_0)$ and $(\tilde{x}_0,\tilde{y}_0)$ in
the attractor, i.e. $y_0=h(x_0)$, $\tilde{y}_0=h(\tilde{x}_0)$.
Their backward orbits up to time $k$ are also contained in the
attractor. We denote them by:
$$
(x_{-k},y_{-k}),\dots,(x_{-1},y_{-1}),(x_0,y_{0})$$ and
$$(\tilde{x}_{-k},\tilde{y}_{-k}),\dots,
(\tilde{x}_{-1},\tilde{y}_{-1}),(\tilde{x}_0,\tilde{y}_{0})\,.
$$
By construction we have that:
\[
\begin{array}{rcl}
x_0&=&f^k(x_{-k})\\ y_0&=&g_{\rho}^{\star
k}(x_{-k},y_{-k})\end{array}
\]
and
\[
\begin{array}{rcl}
\tilde{x}_0&=&f^k(\tilde{x}_{-k})\\
\tilde{y}_0&=&g_{\rho}^{\star k}(\tilde{x}_{-k},\tilde{y}_{-k})\,.
\end{array}
\]
From these equations we can estimate $|y_{0}-\tilde{y}_{0}|$.
Indeed, triangle inequality yields:
\begin{equation}\label{eq1}
\begin{array}{rc}
|y_0\! -\! \tilde{y}_0|
\! \; \leq\! \;
\left|g_{\rho}^{\star k}(x_{-k},y_{-k})\!-\!g_{\rho}^{\star
k}(x_{-k},\tilde{y}_{-k})\right|\!\;\;+ & \\
 \left|g_{\rho}^{\star
k}(x_{-k},\tilde{y}_{-k})\!-\!g_{\rho}^{\star
k}(\tilde{x}_{-k},\tilde{y}_{-k})\right| &
\end{array}
\end{equation}

The first term on the right can be bounded using the contracting
property of $g$:
\begin{equation}\label{eq22}
\vert g_{\rho}^{\star k}(x_{-k},y_{-k})-g_{\rho}^{\star k}(x_{-k},\tilde{y}_{-n})\vert
\leq {\rho}^k |B_y|\, ,
\end{equation}
where $|B_{y}|$ stands for $\mbox{diam}(B_{y})$. The second term
can be bounded by:
\begin{equation}\label{eq3}
\vert g_{\rho}^{\star k}(x_{-k},\tilde{y}_{-k})-
g_{\rho}^{\star k}(\tilde{x}_{-k},\tilde{y}_{-k})\vert\leq L_k
\vert x_{-k}-\tilde{x}_{-k}\vert\,,
\end{equation}
where $L_k$ is the Lipschitz constant of $g_{\rho}^{\star
k}(\cdot,y)$. According to Lemma 16 in the paper~\cite{Afr01}
\begin{equation}\label{eq4}
L_k\leq \frac{\eta}{\gamma_{+}-{\rho}}\,\gamma_{+}^k\;\cdot
\end{equation}
Using assumption (\ref{lip2}) one gets:
\begin{equation}\label{eq5}
\vert x_{-k}-\tilde{x}_{-k}\vert\leq  \gamma_{-}^{k-m}
\vert x_{-m}-\tilde{x}_{-m}\vert\,.
\end{equation}
Putting together these inequalities, one obtains for all $k$:
\begin{equation}\label{eq6}
\vert y_0-\tilde{y}_0|\leq {\rho}^k|B_{y}\vert +
\frac{\eta}{\gamma_{+}-{\rho}}\,\gamma_-^{-m}\,(\gamma_{+}\gamma_{-})^k\vert
x_{-m}-\tilde{x}_{-m}|\,.
\end{equation}
Fix an arbitrarily small $\sigma>0$ and rewrite the first term
in~(\ref{eq6}) as follows
\begin{equation}
{\rho}^k|B_{y}\vert =
({\rho}+\sigma)^k|B_{y}\vert\left(\frac{{\rho}}{{\rho}+\sigma}\right)^k.
\end{equation}
then set
\begin{equation}
\label{set}
({\rho}+\sigma)^k\cong |x_{-m}-\tilde{x}_{-m}|^{\alpha}.
\end{equation}
Therefore using standard logarithmic identity
$$
(\gamma_+\gamma_-)^k\equiv G^{k\log_G (\gamma_+\gamma_-)}
$$
with $G=|x_{-m}-\tilde{x}_{-m}|$, and formula(\ref{set}) one can
write
\begin{equation}
(\gamma_+\gamma_-)^k\cong |x_{-m}-\tilde{x}_{-m}|^{\frac{\alpha\log
(\gamma_+\gamma_-)}{\log ({\rho}+\sigma)}}
\end{equation}
Let $\alpha=\alpha(\sigma)=\left(1-\frac{\log
(\gamma_+\gamma_-)}{\log ({\rho}+\sigma)}\right)^{-1}$.
Then~(\ref{eq6}) implies
\begin{equation}
\begin{array}{cc}
\vert y_0-\tilde{y}_0|\leq \left(\left(\frac{{\rho}}{{\rho}+\sigma}\right)^k
|B_{y}\vert +
\frac{\eta}{\gamma_{+}-{\rho}}\,\gamma_-^{-m}\right) \times &\\  \vert
x_{-m}-\tilde{x}_{-m}|^{\alpha(\sigma)}\,&.
\end{array}
\end{equation}
Thus, if $\vert x_{-m}-\tilde{x}_{-m}|$ is small enough then  $k$
is large enough because of~(\ref{set}). Therefore,
$\left(\frac{{\rho}}{{\rho}+\sigma}\right)^k|B_{y}\vert<\epsilon$,
$\alpha(\sigma)= \alpha_0+\beta(\epsilon)$, and the statement of
Theorem~\ref{theorm2} holds. $\Box$

Hence, we have shown that while the H\"older exponent remains the
same as for $m=0$, the H\"older constant $A_m$ given by formula
(\ref{m-constant}) can be as small as we wish provided that points
on the graph of the function $h$ are close enough. In the numerical
example presented in Sec.~\ref{Sec2} the closeness of driving
trajectories was achieved by selecting the trajectories with the
same symbolic sequence $S_m^i$.

\section{Effects of noise}\label{Sec4}
The studies presented in Sections~\ref{Sec2} and~\ref{Sec3} deal
with the detectability issues of non-differentiable (wrinkled)
synchronization function when data, acquired from drive and
response system, are not contaminated by noise. In a realistic
situation external noise is always present in the data. Taking into
account the complexity of fine structure typical for wrinkled
synchronization functions one may expect that even a very small
noise in the data ruins the detectability of synchronization. We
studied the noise impact using the numerical example considered in
Section~\ref{Sec2}. We examined how the convergence of wrinkled
function $h^{(m)}$ to a polynomial function is effected by external
noise.

The behavior of synchronized systems can be influenced by a noise
in many different ways. For example, stochastic forces, applied to
response and/or drive system, destroy the functional relation
between the systems independently of the complexity of the function
shape. The level of destruction in this case will significantly
depend upon the dynamical properties of the coupled systems.

To be specific we will examine only the case when the
synchronization function exists, but the data representing this
function are contaminated with a noise added to the measurements.
Namely, when one deals with the data $x_n^{(1)}+\xi_n$ and $y_n +
\zeta_n$, where $x_n^{(1)}$ and $y_n$ are generated by equations
(\refmap),(\ref{mapy}), and the independent noisy components
$\xi_n$ and $\zeta_n$ are white Gaussian noise with variances
$\sigma^2_D$ and $\sigma^2_R$, respectively.  In the numerical
analysis we will also assume that symbolic sequences of driving
trajectory are detected correctly.

\begin{figure}
\begin{center}
\includegraphics[width=7.5cm]{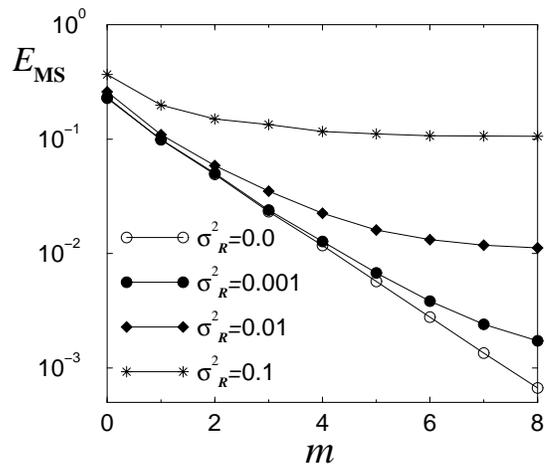}
\end{center}
\caption{The dependence of mean squared error of best polynomial
fitting function for the attractor points $(x_{n-m},y_n)$ on the
mask length $m$ computed for $c=0.7$ and four different values of
variance $\sigma^2_R$ of the noise added to $y_n$ data.}
\label{fig6}
\end{figure}

We found that $\xi_n$ and $\zeta_n$ influence the convergence
properties differently. To illustrate it, consider, first, the
cases when only one source of noise in present. Figure~\ref{fig6}
shows the effect of noise occurred in the measurements of the
response system ($\xi_n=0$ and $\zeta_n\neq0$) for different values
of variance $\sigma^2_R$.
Noise of this type sets a limit on the precision of the function
resolution, see Fig.~\ref{fig6}. The numerical analysis shows that
the limit is $E^*_{MS}\approx
\sigma^2_R$. This result is quite predictable. Indeed, noise in the
response system destroys the function by scattering points along
$y$-variable and makes a thick object (a fuzzy layer)instead of the
graph $h^{(m)}$. The thickness of the layer is characterized by the
level of noise, namely by $\sigma^2_R$. It is clear that, the size
of the thickness (along $y$- variable) does not change under the
transformations applied to the data representing driving
trajectory. Since our method is not designed to locate the function
inside this fuzzy layer, we cannot expect the accuracy (in terms of
$E_{RM}$) be better than $\sigma^2_R$.

\begin{figure}
\begin{center}
\includegraphics[width=7.5cm]{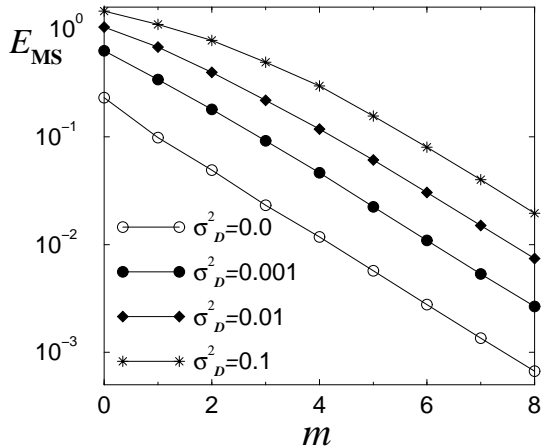}
\end{center}
\caption{The dependence of $E_{MS}$ on $m$ computed for $c=0.7$ and
four different values of variance $\sigma^2_D$ of the noise added
to $x_n^{(1)}$ data.}
\label{fig7}
\end{figure}

When noise occur in the measurements of driving variable
($\xi_n\neq0$ and $\zeta_n=0$) the conversion process has a
different dynamics, see Fig.~\ref{fig7}. Now the graph of the
function in Fig.~\ref{fig1} is transformed into a thick layer due
to scattering of data points along variable $x$. However, as it
follows from Fig.~\ref{fig7} that thickness does not limit the
precision of function evaluation. This effect can be understood
from the considerations presented in Fig.~\ref{fig1} and
Fig.~\ref{fig5}. Indeed, the increase of trajectory length in the
analysis of the function is equivalent to rescaling of the function
argument, while the overall shape of the function does not change,
see Fig.~\ref{fig5}. While the interval of the argument values
increases with the trajectory length as $2^m$ the $x$-size of the
think layer remains unchange. As the result the increase of $m$, in
this case, improves the precision function detection.

\begin{figure}
\begin{center}
\includegraphics[width=7.5cm]{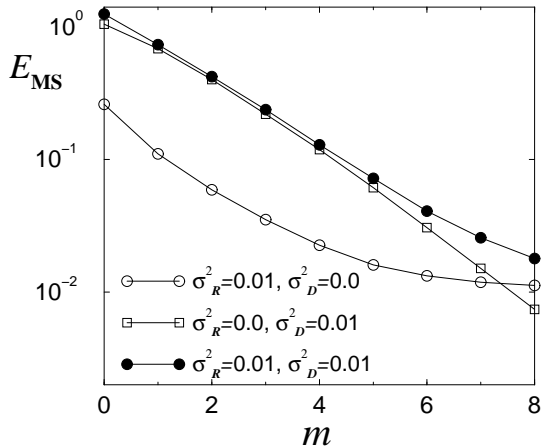}
\end{center}
\caption{The dependence of mean squared error of best polynomial
fitting function for the attractor points $(x_{n-m},y_n)$ on the
trajectory length $m$ computed for three different n.}
\label{fig8}
\end{figure}

The effects induced by additive noise in the data is summarized in
Fig.~\ref{fig8} where the dependence of $E_{MS}$ versus $m$ is
shown for three different situation of noise ($\xi_n=0$,
$\zeta_n\neq0$), ($\xi_n\neq0$, $\zeta_n=0$) and ($\xi_n\neq0$,
$\zeta_n\neq0$). These plots are computed for the coupling strength
$c=0.7$. The results indicate that if the level of noise in the
data is small a non-differentiable function is detectable with some
accuracy limited by the noise variance.

In this analysis we assumed that symbolic sequence representing the
driving trajectory is detected correctly. We expect that when the
symbolic sequence contains errors, then, the function can be
severely damaged and, the improvement of function detection can
fail. We believe that the use of a noise reduction technique on the
data before the computation of $E_{MS}$ versus $m$ plots can be
very beneficial in this case.

\section{Conclusions}

A simple numerical example and rigorous theoretical analysis show
that, despite complex shapes of non-differentiable synchronization
functions, the existence of such a function can be detectable in
practical situation. This can be achieved by considering this
function not as a function of the current state of driving system,
but as a function of the state, in which the driving system has
been a few steps ago. Thanks to the contracting properties caused
the dissipation in the driving system the nearby trajectories
disperse far away from each other in the previous times. This
effect can be used as a "magnifying glass" in the detection of
non-differentiable synchronization function that contains multiple
wrinkles and cusps. It is shown that, although the "magnified"
function in this analysis remains non-differentiable, the amplitude
of these wrinkles and cusps gets smaller as the delay increases.

The results presented in this paper are in agreement with the
recent study on detecting of generalized synchrony made by He,
Zheng and Stone~\cite{He03}. Their study based on a different
technique that involves the analysis of drive and response
trajectories in the embedded phase spaces and takes into account
$p$ preimages for each trajectory point. The use of the preimages
in this case also acts as a "magnifying glass" in detecting of the
wrinkled synchronization function.

In the example considered in Section~\ref{Sec2}, the
synchronization function depends only on the $x^{(1)}$--variable,
which is the coordinate that always represents a stable direction
in ${\bf x}$. Stable and unstable direction in the ${\bf x}$-space
in this driving system are fixed and do not dependent on ${\bf x}$.
Therefore, the differential of this map is a constant matrix. In a
more general situation, this is not the case, and the
synchronization function must depend on both "stable" and
"unstable" coordinates. Nevertheless, it is possible to understand
(although it is not so simple to prove) that the dependence on the
unstable coordinate is non-essential in the hyperbolic situations.
The simplest way to be convinced is to remember that for hyperbolic
attractor there exists a local H\"older-continuous change of
variables such that in new variables the stable and unstable
directions are along the coordinate lines (planes) and the
situation becomes very similar to the example considered.

We examined the influence of external noise on the function
improvement. We found that noise in the data acquired from the
response system sets a limit for the accuracy of the function
approximation and, as the result, after some critical value of $m$
the further increase of delay becomes useless.

To conclude we would like to emphasize that, although we apply our
study to the theory of chaos synchronization, the data analysis
method and theory developed here can be useful for other
applications. Such applications include prediction of chaotic
dynamical behavior in time and space and other studies associated
with various types of prediction. The use of hybrid,
"continuous-symbolic" representation of the chaotic trajectories
enables one to take into account additional information about the
trajectory in a compact way.

\section{Acknowledgments}
The authors are grateful to A. Cordonet, J. Urias, L.S. Tsimring,
H.D.I. Abarbanel, and M.I. Rabinovich for stimulating discussions.
This work was supported in part by a grant from the University of
California Institute for Mexico and the United States (UC MEXUS)
and the Consejo Nacional de Ciencia y Tecnologia de M\'{e}xico
(CONACYT). N.R. was sponsored in part by U.S. Department of Energy
(grant DE-FG03-95ER14516) and the U.S. Army Research Office (MURI
grant DAAG55-98-1-0269). V.A was partially supported by CONACyT
grant 36445-E.


\begin{thebibliography}{99}

\bibitem{Glass} L. Glass and M.C. Mackey, {\em From clocks to
chaos: the rhythms of life} (Princeton, N.J. : Princeton University
Press, 1988), 248.

\bibitem{Minorsky} N. Minorsky, Nonlinear oscillations, [Huntington,
N.Y., R. E. Krieger Pub. Co.], 1974 [c1962], 714.

\bibitem{Blekhman} I.I. Blekhman, Synchronization in Science and
Technology [Amer Society of Mechanical Engineers], 1988, 255.

\bibitem{Pikovsky} A. Pikovsky, M. Rosenblum, and J. Kurths
Synchronization: A Universal Concept in Nonlinear Science
[Cambridge University Press], 2002, 500.

\bibitem{Mosekilde} E. Mosekilde, Yu. Maistrenko, and D. Postnov, Chaotic Synchronization: Applications to Living
Systems [World Scientific Publishing Co., Inc.], 2002, 440.

\bibitem{Boccaletti} S. Boccaletti, J. Kurths, G. Osipov, D.L. Valladares,
and C.S. Zhou, Physics Reports {\bf 366}, 1 (2002).

\bibitem{AVR86} V. Afraimovich, N.N. Verichev and M.I. Rabinovich,
Radiophys. Quant. Electr. {\bf 29}, 747 (1986).

\bibitem{FY84}H. Fujisaka and T. Yamada. Prog. Theor. Phys. {\bf 69},
32 (1984)

\bibitem{Pecora90} L.M. Pecora and T.L. Carroll. Phys. Rev. Lett.
{\bf 64} (1990), 821-824.

\bibitem{Rulkov95} N.F. Rulkov, M.M. Sushchik, L.S. Tsimring, and
H.D.I. Abarbanel, Phys. Rev E {\bf 51}, 980 (1995).

\bibitem{Pecora95} L.M. Pecora, T.L. Carroll, and J.F. Heagy,
Phys. Rev E {\bf 52}, 3420 (1995).

\bibitem{Parlitz96} L. Kocarev and U. Parlitz, Phys. Rev. Lett.
{\bf 76}, 1816 (1996).

\bibitem{Rulkov2001} N.F. Rulkov, V.S. Afraimovich, C.T. Lewis,
J.-R. Chazottes, and A. Cordonet, Phys. Rev E {\bf 64}, 016217
(2001).

\bibitem{Pikovsky97}  A. Pikovsky, M. Zaks, M. Rosenblum, G. Osipov, and
J. Kurths, Chaos {\bf 7}, 680 (1997)

\bibitem{Zaks99} M.A. Zaks, E.H. Park, M.G. Rosenblum, and J Kurths,
Phys. Rev. Lett {\bf 82}, 4228 (1999).

\bibitem{VDP} B. van der Pol, Philos. Mag., {\bf 3}, 65 (1927).

\bibitem{Andronov} A.A. Andronov and A.A Witt, Atch. Elektrotech {\bf 16},
280 (1930).

\bibitem{Abarbanel96} H.D.I. Abarbanel, N.F. Rulkov and M.M. Sushchik,
Phys. Rev. E {\bf 53}, 4528 (1996).

\bibitem{Stark97} J. Stark, Physica D {\bf 10}, 163 (1997).

\bibitem{Hunt97} B.R. Hunt, E. Ott and J.A. Yorke, Phys Rev. E
{\bf 55}, 4029 (1997);

\bibitem{Josic98} K. Josi\'{c}, Phys. Rev. Lett.
{\bf 80}, 3053 (1998).

\bibitem{Stark99} J. Stark, Ergod. Theory Dyn. Syst. {\bf 19}, 155 (1999).

\bibitem{Afr01} V. Afraimovich, J.-R. Chazottes, and
A. Cordonet, Discrete and Continuous Dyn. Systems: Ser. B {\bf 1},
421 (2001).

\bibitem{So02} P. So, E. Barreto, K. Josi\'{c}, E. Sander, S.J. Schiff,
Phys. Rev. E {\bf 65} 046225 (2002).

\bibitem{Barreto} E. Barreto, K. Josi\'{c}, C. Morales, E. Sander, and
P. So, Chaos {\bf 13} 151 (2003).

\bibitem{ACR02} V. Afraimovich, A. Cordonet, and N.F. Rulkov, Phys. Rev. E
{\bf 66} 016208 (2002).

\bibitem{He03} D. He, Z. Zheng, and L. Stone, Phys. Rev. E
{\bf 67} 026223 (2003).

\end{thebibliography}
\end{document}